\documentclass[aps,pra,twocolumn,superscriptaddress,showpacs,showkeys,10pt]{revtex4-1}

\def\ttitle{Pressure-Driven Evaporative Cooling in Atom Guides}

\usepackage{graphicx}
\usepackage{amsmath}
\usepackage{color}
\usepackage{eepic}
\usepackage{amssymb}
\usepackage{amsxtra}           
\usepackage{xspace}            
\usepackage{ifthen}            
\usepackage{shortcuts}
\usepackage{symb}
\usepackage[refpages]{gloss}

\usepackage[pdfpagelabels=true]{hyperref}

\hypersetup{
    naturalnames=true,
    colorlinks=true,
    linkcolor=blue,
    pdfpagemode=UseNone,
    pdfstartview=FitH,
    pdftitle={\ttitle},
    pdfauthor={Spencer E. Olson},
    pdfsubject={Evaporative cooling of magnetically guided cold atoms},
    pdfkeywords={Evaporation, Cold atoms, Atom guide, DSMC },
}
\usepackage{hypernat}

\newboolean{publish}
\setboolean{publish}{true}

\newgloss[nolink]{symb}{.sym}{List Of Symbols}{glslookup}

\newcommand{\PRsection}[1]{}

\begin{document}

\title{\ttitle}

\author{Spencer E. Olson}
\affiliation{
    Air Force Research Laboratory, Space Vehicles Directorate,  \\
    3550 Aberdeen Ave. SE, Kirtland Air Force Base, NM 87117-5776
}

\author{Georg Raithel}
\affiliation{
    Physics Department, University of Michigan\\
    2477 Randall Lab, 450 Church Street, Ann Arbor, Michigan 48109-1120
}

\author{Andrew J. Christlieb}
\affiliation{
    Mathematics Department, Michigan State University\\
    D304 Wells Hall, East Lansing, Michigan 48824-1027
}

\date{\today}

\begin{abstract}
We study steady-state evaporation in an atom guide via Monte Carlo simulations.
The evaporation surface follows a specific profile as a function of longitudinal
guide location.  We demonstrate that the choice of evaporation profile
significantly impacts the performance of the evaporation.  Our simulations also
demonstrate a significant performance boost in the evaporation when using a
longitudinally compressed guide.  We show that for a purely pressure-driven atom
beam, it should be possible to reach degeneracy within a $0.5~\m$ guide for
experimentally feasible, albeit challenging, loading conditions.
\end{abstract}



\maketitle


\printgloss[symb]{symbols} 
\ifthenelse{\boolean{publish}}{ 
\lookupitem{B0}{\ensuremath{B_{0}}}{\ensuremath{B_{0}}}{}{Small (longitudinal)
  bias of trap to prevent non-adiabatic Majorana spin flips.}

\lookupitem{B}{\ensuremath{\vec{\bf B}\!\left(\vec{\bf
  x}\right)}}{\ensuremath{\vec{\bf B}\!\left(\vec{\bf x}\right)}}{}{Magnetic
  field vector}

\lookupitem{F}{\ensuremath{F}}{\ensuremath{F}}{}{Hyperfine level quantum
  number}

\lookupitem{FN}{\ensuremath{F_{\rm N}}}{\ensuremath{F_{\rm N}}}{}{Size of
  representative particle (in units of number of real particles)}

\lookupitem{evap_func}{\ensuremath{\mathfrak{F}\left(z\right)}}{\ensuremath{\mathfrak{F}\left(z\right)}}{}{Phase-space
  distribution function (see Boltzmann Eq.)}

\lookupitem{gamma_coll}{\ensuremath{\gamma_{\rm coll}}}{\ensuremath{\gamma_{\rm
  coll}}}{}{Elastic scattering rate}

\lookupitem{K_B}{\ensuremath{k_{\rm B}}}{\ensuremath{k_{\rm B}}}{}{Boltzmann
  constant}

\lookupitem{l_th}{\ensuremath{\lambda_{\rm th}}}{\ensuremath{\lambda_{\rm
  th}}}{}{Thermal de~Broglie wavelength}

\lookupitem{mass}{\ensuremath{M}}{\ensuremath{M}}{}{Mass}

\lookupitem{mF}{\ensuremath{m_{\rm F}}}{\ensuremath{m_{\rm F}}}{}{Hyperfine
  sublevel quantum number}

\lookupitem{mu_B}{\ensuremath{\mu_{\rm B}}}{\ensuremath{\mu_{\rm B}}}{}{Bohr
  magneton}

\lookupitem{nl3}{\ensuremath{n \lambda_{\rm th}^{3}}}{\ensuremath{n
  \lambda_{\rm th}^{3}}}{}{Phase space density}

\lookupitem{scatT}{\ensuremath{\sigma_{\rm T}}}{\ensuremath{\sigma_{\rm
  T}}}{}{Total elastic scattering cross section}

\lookupitem{vlong}{\ensuremath{v_{\parallel}}}{\ensuremath{v_{\parallel}}}{}{Velocity
  in longitudinal direction}
 }{}

\PRsection{Introduction}

The past several years have seen significant research to develop a
continuous-wave (CW) atom laser~\cite{Robins:2013:alp}.  Analogous to the
impacts the CW optical laser had on precision measurements, a CW atom laser is
expected to impact precision atom-based metrology via longer coherence lengths
and greater continuity of temporal measurement coverage.  A CW laser is
identified by a continuously loaded and leaked macroscopic occupation of a
quantum wave-function.  The macroscopic-occupation state for an atom laser is
a Bose-Einstein Condensate (BEC), where ultra-cold atoms can condense into
the single ground state of a reservoir trap.  Thus, in order for a CW atom laser to be established,
atoms must be continuously cooled to sub-microkelvin temperatures, transported,
and loaded into the BEC.  

One of the primary tools for obtaining sub-microkelvin temperatures in atomic
systems entails the use of forced evaporative cooling~\cite{Ketterle1996a}.
Evaporative cooling consists of removing the most energetic particles of a
system, thereby lowering the total energy of the system.  With forced
evaporative cooling, the evaporation threshold is strategically lowered in order
to maintain the energy-removal rate.  As thermal energy is removed from an ensemble
of atoms, the first-order coherence within the sample increases, often
increasing the signal-to-noise ratio and precision of atomic measurements.
This technique has been key in the forming of the Bose-Einstein condensate state
of matter and has proven useful in many other cold-atom applications..

\begin{figure}[htb]
  \centerline{
    \includegraphics[width=\columnwidth]{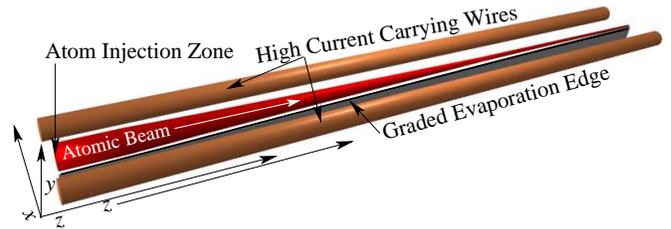}
  }
    \caption[Conceptual figure] {
        \label{fig:concept}
        Continuous evaporation is distributed in space and maintained in time by continually narrowing an
        effective evaporative surface in towards the center of the atom beam.
    }
\end{figure}

Some approaches for developing a CW atom laser attempt to establish a
steady-state evaporative cooling process along the longitudinal direction of a
guided cold atom beam.
In this manner, an atomic beam is transferred into and transported by
a magnetic guide~\cite{Olson:cpe2006} where the beam is cooled
as it travels through the guide~\cite{Lahaye2005a}.  Beam
temperature is lowered using an evaporation threshold that varies as a function
of longitudinal position.  Fig.~\ref{fig:concept} shows a conceptual picture of
distributed evaporative cooling.  An atomic beam, injected at $z=0$,
travels down the length of a magnetic guide.  During this travel, an evaporation
edge gradually removes more and more atoms from the beam.  By maintaining
collisionality throughout the guide, the beam is locally thermally equilibrated at all
longitudinal positions, resulting in a continual decrease of the beam
temperature~\cite{Mandonnet:eca2000,Lahaye:2006:kec}.

The goal of this paper is to demonstrate that a realistic and efficient
strategy exists for establishing steady-state forced evaporation along the
length of an atom beam.
While the use of a supersonic beam of cold
atoms is usually justified in order to inhibit thermal shortcuts between
longitudinally distant atoms, we demonstrate that a pressure-driven flow has unique advantage for
establishing steady-state cooling:  not only are thermal shortcuts eliminated
via collisionally viscous flow, but evaporative processes can be tuned within
significantly shorter distances.  
We study a basic set of longitudinally varying
evaporation surfaces that might be applied in an experiment.  Simulation
results are presented to depict the relative performance among the various
strategies, and a clear relative optimum is demonstrated.

By employing a method known as Direct Simulation Monte Carlo
(DSMC)~\cite{Bird:mgd1994}, we simulate the evaporative
cooling process in the magnetic guide.  Simulations offer a reliable preview
of the performance of a particular evaporation design.
The DSMC method has been used successfully to simulate cold-atom collision
processes such as evaporative cooling~\cite{Wu:dsec1996,Mandonnet:eca2000} and
s/d-wave collisions statistics~\cite{Wade:dsmc:2011}.
In this work, we use a new gridless DSMC algorithm that provides uniform
accuracy independent of density modulations.  Our algorithm, the advantages
thereof, and thorough testing are more thoroughly discussed in Ref.~\cite{Olson:gdsmc:2008}.
Simulations of the atom guide presented here were done using a parallel
algorithm discussed in Ref.~\cite{Olson:pid:2010}.


\PRsection{Key Concepts}


There are two key ingredients for establishing an efficient temperature gradient
along the length of an atom guide using evaporative cooling:  rethermalization
and thermal isolation.  Rethermalization occurs as a locally disturbed gas
undergoes collisions to reach local thermal equilibrium.  Thermal isolation in a
guide, on the other hand, pertains to disallowing direct energy exchange between longitudinally
distant groups of atoms.  

\begin{figure}[htb]
  \centerline{ \includegraphics[width=\columnwidth]{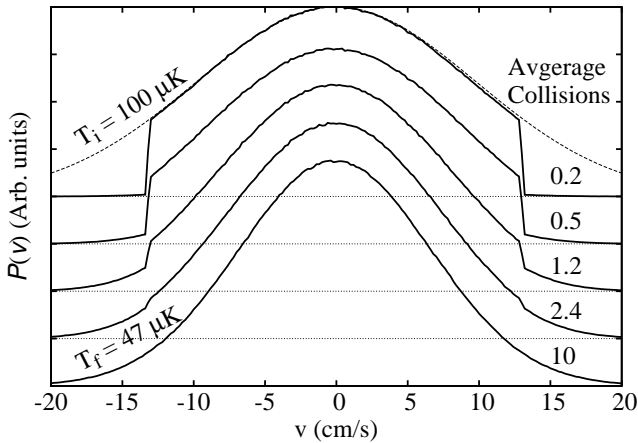} }
  \caption[Rethermalization for an initially velocity-truncated distribution] {
    \label{fig:retherm}
    The velocity distribution of a $100~\uK$ sample of $^{87}$Rb atoms is
    truncated and allowed to rethermalize via collisions.  Approximately $20\%$,
    of the initial atom population, is removed as the wings of the thermal
    distribution are truncated.  After roughly $10$ collisions per atom, the
    velocity distribution errs, on average, less than $(1/100)\%$ from a thermal
    distribution with $T = 47~\uK$. 
  }
\end{figure}

As an example of rethermalization, consider a uniform gas of
$^{87}$Rb atoms with initial conditions given by $n_{\rm i} =
\sci{1.25}{8}~\cm^{-3}$ and $T_{\rm i} = 100~\uK$, where $n_{\rm i}$ and $T_{\rm
i}$ are the initial number density and temperature of the sample respectively.
The velocity distribution of this gas can be represented by a normal
distribution as shown in the top dashed curve in Fig.~\ref{fig:retherm}.  
A disturbance is created by suddenly removing the $20\%$ most energetic
particles such that the velocity distribution is truncated as approximately
represented by the highest solid curve in Fig.~\ref{fig:retherm}.  As shown in
Fig.~\ref{fig:retherm}, the rethermalization process begins immediately with the
first collisions after the truncation.  Each collision quickly brings the
velocity distribution of the gas closer and closer to a Boltmann distribution
until the sample reaches thermal equilibrium.  Because energy was removed via
the truncation, the resulting Boltzmann distribution is narrower and more
peaked, representing a colder collection of atoms.  Depending on the final error
admissible, Fig.~\ref{fig:retherm} shows that the rethermalization time is in
the range of $2$--$10 \symb{gamma_coll}^{-1}$ where $\symb{gamma_coll}$ is the
average collision rate~\cite{Wu:dsec1996,Snoke:pdb1989,Monroe:mcs1993}.  

Thermal isolation in the guide direction, $z$, is critical to ensure that hotter,
upstream atoms cannot reach, collide with, and heat downstream portions of the
atomic beam.  Problematic longitudinal heat conduction arises from (1) low stream
densities that result in collisionless flow or (2) high-angular-momentum
trajectories of beam atoms through the guide~\cite{Meppelink:ehf:2009}.
Avoiding heat conduction is made possible by first maintaining
collisionality and second, removing atoms that reach large guiding radii.
High collisionality ensures that atoms exchange momentum only
with other atoms of similar kinetic energy in nearly the same portion of the
beam.  By choosing an injection velocity $\symb{vlong}_0$ on the same order as the initial
thermal velocity, flow is pressure driven instead of being primarily due to the
inertia of the incident flow.  Evaporative removal of all atoms outside of a
$z$-dependent, critical radius ensures that high-angular-momentum states are
disallowed.

\PRsection{Evaporation Function}

\begin{figure}[htb]
    \centerline{ \includegraphics[width=\columnwidth]{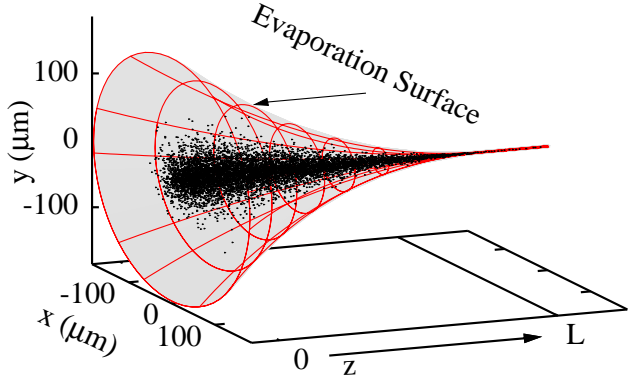} }
    \caption[Picture of atoms in the forced evaporation guide] {
        \label{fig:atoms}
        Snapshot of the guided atom beam under the influence of forced
        evaporative cooling.  The 3D surface indicates the location of the
        evaporation surface. 
    }
\end{figure}

Assuming that a particular temperature gradient can be established, the goal
becomes to identify the type of gradient which results in the highest
final phase-space density \symb{nl3} where $n$ is the number density and
\symb{l_th} is the thermal de~Broglie wavelength given by
\[
    \symbi{l_th} =  \sqrt{ \frac{2\pi\hbar^{2}} {\symbi{mass} \symbi{K_B} T} }
    \; .
\]
Neglecting the longitudinal potential as a degree of freedom, each particle in
the system has an average energy equal to $5\symb{K_B}T/2$.  For simplicity, we
assume that the edge
of the near-thermal spatial distribution is at a transverse potential energy of
$15\symb{K_B}T/2$.  The exact value of this assumption is arbitrary but must
result in a nearly total inclusion of the number and kinetic-energy distributions.  For
the choice of $15\symb{K_B}T/2$, $99.9\%$ of the distribution with $99.5\%$ of
the kinetic energy lies within this range.  By removing atoms that reach
$U(\vec{\bf x}) = 15\symb{K_B}T/2 + \Delta$, where $\Delta$ is the minimum
energy in the center of the trap given by $\Delta = g_{F} m_{F} \mu_{B}
\symb{B0}$, we can thus force the guide to support a temperature no greater than
$T$.  To establish a particular temperature gradient through the guide, we
define the evaporation threshold function \symb{evap_func} as the potential
energy $U(\vec{\bf x})$ at which atoms are removed at a longitudinal location
$z$ in the guide.  This results in an atomic distribution that increasingly
narrows as the beam progresses down the guide.  Fig.~\ref{fig:atoms}
shows such an atom distribution and the corresponding evaporation threshold
surface that causes this narrowing,
where $L$ is the length of the guide over which the evaporation threshold
changes.

\begin{figure}[htb]
    \centerline{ \includegraphics[width=\columnwidth]{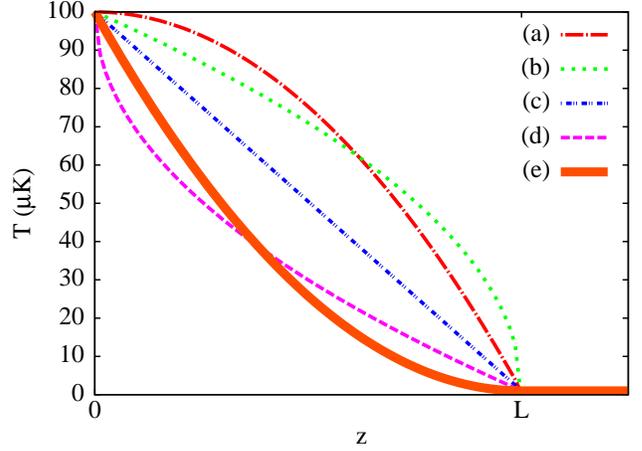} }
    \caption[Evaporation strategies of interest] {
        \label{fig:evap-types}
        Evaporation strategies of interest for this paper.  The curves here
        depict the radial position $\rho$ at which atoms are removed from the
        system.  By lowering the barrier in the forward direction, a forced
        evaporative cooling is imposed.  
        The curves (a)--(e) represent the lines of \symb{evap_func}
        (Eq.~\ref{eq:evap_func}), in order respectively.  
        For $z \leq 0$ and $z \geq L$,
        $\symb{evap_func}$ is held constant. 
        These different strategies are to provide a set of
        basic types of evaporation surfaces that might be used experimentally. 
    }
\end{figure}

To evaluate the effect of the evaporation threshold function
$\symb{evap_func}$ on the final phase-space density $\symb{nl3}$, we explore a
(somewhat arbitrary) basic set given by: 
\begin{equation}
    \label{eq:evap_func}
\symb{evap_func} \sim
    \left\{
        \begin{array}{rll}
            T_{\rm f} + \left(T_{\rm i} - T_{\rm f}\right) & \cdot~
                \left[1\right. - &\left(z/L\right)]^{1/2}, \\
            T_{\rm i} + \left(T_{\rm f} - T_{\rm i}\right) & \cdot 
                     &\left(z/L\right)^{2}, \\
            T_{\rm i} + \left(T_{\rm f} - T_{\rm i}\right) & \cdot
                     &\left(z/L\right), \\
            T_{\rm i} + \left(T_{\rm f} - T_{\rm i}\right) & \cdot
                     &\left(z/L\right)^{1/2}, \\
            T_{\rm f} + \left(T_{\rm i} - T_{\rm f}\right) & \cdot~
                \left[1\right. - &\left(z/L\right)]^{2}
        \end{array}
    \right\}
    \,,\, 0 \leq z \leq L
\end{equation}
Each of the basic $\symb{evap_func}$ functions in Eq.~\ref{eq:evap_func}, shown in
Fig.~\ref{fig:evap-types}, serves to evaluate a particular strategy for
evaporative cooling.  Each represents a different level of compromise between the
need to establish thermal isolation (by removing high energy atoms earlier)
and the need to achieve a high density in the downstream portions of the
guide.  

As the atoms travel down the guide, the evaporation process
depletes the atom number and can eventually stall (on the timescale of the
guide traversal) as the collision rate decreases.  To prevent the forced
evaporation from stalling, the atoms are magnetically squeezed to enhance the collision rate.
In a two-wire magnetic guide, this is done by simply decreasing the separation
between the two wires as a function of $z$.  The computed magnetic field used
for this work corresponds to two parallel currents ($I = 150~\amps$) with a separation of $5.175~\mm$ at $z=0$ linearly
decreasing to $4.175~\mm$ at $z = 50~\cm$.  This separation results in
gradients similar to the apparatus in Ref.~\cite{Olson:cpe2006}.  
Compression enhances the density and thus helps to maintain the collision rate
as the atom number diminishes through the evaporative cooling process. 
Compression as a function of longitudinal position in the guide is analogous to temporal
compression performed in standard BEC formation~\cite{Hess:ecm1986,
Monroe:mcs1993}.
It should be noted that the work done in Ref.~\cite{Mandonnet:eca2000}
simulated evaporation in a non-compressed magnetic guide.

After forced evaporation ends at $z = L = 40~\cm$, the atoms continue until
they reach an elastically reflecting barrier at $z = 50~\cm$.  As described in
~\cite{Olson:cpe2006}, this wall could be created by a blue-detuned sheet of
light, such that a three dimensional trap is established wherein a condensate
can form.

\PRsection{Configuration}

To simulate the atom guide, $^{87}$Rb atoms are injected at $z=0$ into an
initially empty guide at a rate of $\sci{3}{9}~\s^{-1} \Delta t$ per timestep
$\Delta t$ with an average stream velocity of $\symb{vlong}=0$ and a
temperature $T = 100~\uK$.  For these simulations, $L = 40~\cm$.  As allowed by the Boltzmann equation, each
simulated particle is scaled to represent $\symb{FN}\ge1$ rubidium atoms, such
that each simulated particle has a cross section $\symb{FN} \symb{scatT}$ where
\symb{scatT} is the total cross section of a single particle.
Scaled representative particles decrease the computational burden and are
typical in gas dynamics simulations.  The figure of merit for choosing
\symb{FN} is the ratio of the average distance between colliding particles to
the mean free path~\cite{Olson:gdsmc:2008}.  A large \symb{FN} results in
higher values of this ratio.  If this ratio is too large ($\gtrsim 1$), the
collisionality of the simulation is significantly diminished.  For this work,
\symb{FN} was chosen as high as possible ($\sci{5}{3}$) without significantly
diminishing the collision processes.  It should be noted that a decrease in
\symb{FN} will only result in a more collisional system and a larger
temperature gradient along the length of the guide.  These simulations
therefore represent a lower bound of the evaporative-cooling performance.  

By using a very low input stream velocity, it is expected that the pressure
driven flow will allow a very short guide that is still able to maintain a
strong temperature gradient.  The input source is assumed to be in the
$\left|\symb{F}=2,\symb{mF}=2\right>$ stretched state.  Thus, atoms are guided
by a potential $U(\vec{\bf x})$ given by
\begin{equation}
    \label{eq:U}
    U(\vec{\bf x}) = m g \vec{\bf x}\cdot \hat{\bf y}
      + \symb{mu_B}
        \left|\vec{\bf \symb{B0}} + \symb{B}\right|
\end{equation}
where $\symb{B}$ is the magnetic field resulting from two
parallel wire currents and $\symb{B0}$ is a small longitudinal bias
(0.5~\G) to prevent non-adiabatic Majorana spin flips.
The injected atom stream is mode-matched to the guide according to the
distribution $P(\vec{\bf x},\vec{\bf v}) |_{z=0}$ given by
\begin{equation}
\label{eq:P}
P(\vec{\bf x},\vec{\bf v}) |_{z=0} \propto
    \exp
        \left[
           -\frac{\symb{mass} \left| \vec{\bf v} 
                              \right|^{2} }
                 { 2 \symb{K_B} T}
           -\frac{ U(\vec{\bf x}) |_{z=0} }
                 { \symb{K_B} T}
        \right]\, .
\end{equation}
%
%
%
%
Each timestep of the simulation is broken into two major sub-steps:
collisionless motion followed by momentum exchanging collisions within
nearest-neighbor atom groups, as described in Ref.~\cite{Olson:gdsmc:2008}.

\PRsection{Results}

\begin{figure}
    \centerline{ \includegraphics[width=\columnwidth]{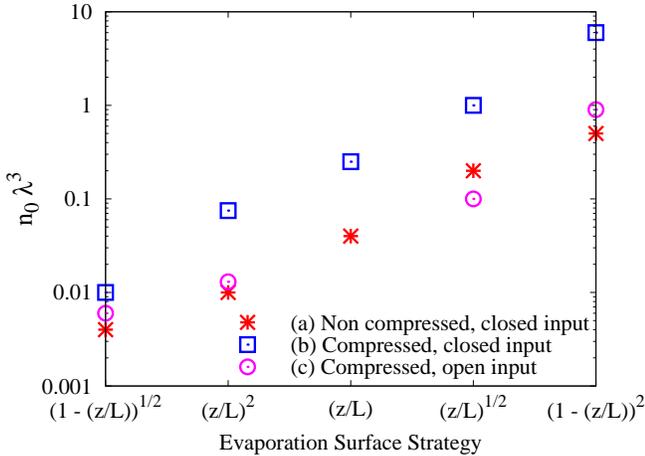} }
    \caption[\symb{nl3} of the different evaporation strategies] {
        \label{fig:results-nl3}
        Final phase space density $\symb{nl3}$ of the different evaporation
        strategies. 
        (a)  Non-compressed guide has a constant magnetic field
        gradient of $750~\G/\cm$.  A reflecting barrier is also placed at $-4~\cm$.
        (b)  Compressed guide begins with a gradient of $750~\G/\cm$ and ends with
        $1500~\G/\cm$.  The reflecting barrier at $-4~\cm$ is also present here.
        (c)  Compressed guide ($750~\G/\cm \rightarrow 1500~\G/\cm$) without
        reflective barrier at $-4~\cm$.  
    }
\end{figure}

To quantify the results of the simulated evaporative cooling, we examine the
increase in \symb{nl3} over the length of the guide.  For all cases,
$\symb{nl3} \sim 10^{-4}$ at $z=0$.  Fig.~\ref{fig:results-nl3} compares the
value of \symb{nl3} at $z=L$ for each instance of \symb{evap_func} in
Eq.~\ref{eq:evap_func}.  For the results shown in Fig.~\ref{fig:results-nl3},
it is clear that the performance depends greatly on the form of
\symb{evap_func}, with $\symb{evap_func}\sim(1- (z/L))^2$ being the best
candidate and $\symb{evap_func}\sim(1-(z/L))^{1/2}$ being the least promising.

Fig.~\ref{fig:results-nl3} also compares the following cases for each type of
\symb{evap_func}:  (a) a non-compressed guide with a wall at $-4~\cm$, (b) a
compressed guide with a wall at $-4~\cm$, and (c) a compressed guide with {\it
no} wall at $-4~\cm$.  For cases (a-b), the reflecting barrier at $-4~\cm$
boosts the density throughout the guide and provides a best case scenario for
comparison.  This comparison shows that the compressed guide does indeed
result in a higher final value of \symb{nl3}.   

\PRsection{Conclusion and Discussion}

Our simulations show that an optimum strategy does exist for cooling atoms in a
guide.  By appropriately choosing the evaporation threshold function
$\symb{evap_func}$, it should be possible to achieve degeneracy even with short
atom guides.  Furthermore, as stated earlier, these simulations are expected to
show only increased thermal isolation in the $z$ direction and hence
greater evaporative-cooling performance as the representative particle size
$\symb{FN} \rightarrow 1$.   
In addition, as $\symb{nl3} \rightarrow 1$, Bose statistics are
expected to accelerate the condensation process of atoms into the ground state
of the atom guide.  We therefore conclude that a more accurate model of the physics
would predict even higher evaporative cooling efficiency.

For high-precision metrology, one might use the evaporation profiles
described here to obtain a steady-state (stationary in time) BEC at the end of
the guide.  
For this, it is apparent that additional controls must be
introduced for reducing the forward stream velocity.   
The most practical control variable for removing the
longitudinal energy is the tilt applied to the guide, such that atoms are
forced to climb a gravitational potential.  Other methods of removing
longitudinal energy could include moving
magnetic~\cite{Reinaudi:2006:mmm,Bethlem:2008:ghd,imhof2013chip} or perhaps optical potentials.

\begin{acknowledgments}
This work was supported by the Naval Research Laboratory, Department of
Defense High Performance Computing Modernization Program, and the Army Research
Office (Project number 42791-PH).  A.~J. Christlieb was supported from AFOSR grants
FA9550-11-1-0281, FA9550-12-1-0343.
\end{acknowledgments}

\ifthenelse{\boolean{publish}}%
{

}{
  \bibliography{dsmc-guide}
}

\end{document}